\documentclass[aps,prl,epsfigure,twocolumn,
]{revtex4}
\usepackage{graphicx}
\usepackage{amsmath}    
\usepackage{amsfonts}   
\usepackage{epsfig}
\usepackage{color}

\newcommand{\miniket}[1]{\vert#1\rangle}

\newcommand{\minibra}[1]{\langle#1\vert}

\begin{document}

\title{Mimicking the probability distribution of a two-dimensional Grover walk with a single-qubit coin}

\author{C. Di Franco$^1$, M. Mc Gettrick$^2$ and Th. Busch$^1$}

\affiliation{$^1$ Department of Physics, University College Cork, Cork, Republic of Ireland\\
$^2$ The De Br\'un Centre for Computational Algebra, School of Mathematics, The National University of Ireland, Galway, Republic of Ireland}

\begin{abstract}
The non-localized case of the spatial density probability of the two-dimensional Grover walk can be obtained using only a two-dimensional coin space and a quantum walk in alternate directions. This significantly reduces the resources necessary for its feasible experimental realization. We present a formal proof of this correspondence and analyze the behavior of the {\sl coin-position entanglement} as well as the {\sl x-y spatial entanglement} in our scheme with respect to the Grover one. Our scheme allows us to entangle the two orthogonal directions of the walk more efficiently.
\end{abstract}



\maketitle

The concept of a random walk, as the mathematical description of a sequence of random steps, has a large number of applications in different fields, ranging from physics to computer science and from economics to biology~\cite{RWGeneral}. Its analogue in quantum theory is the so-called quantum walk which, due to the unitarity requirement, brings with it certain changes~\cite{Aharonov:93,Kempe:03}: while in a classical random walk the position of the particle (walker) is described by a probability distribution, in a quantum walk the particle is in a superposition of quantum states. It has been shown that any quantum circuit can be efficiently simulated using quantum walks~\cite{Childs:09} and interesting proposals to implement quantum walks in different physical systems have been put forward~\cite{qwimplementations,Eckert:05}. An equivalence between the dynamics of quantum walks and quantum lattice gas models~\cite{feynman} has also been found.

One of the key properties of quantum walks is their ability to evolve disentangled states into entangled ones and to efficiently generate entanglement in experimentally feasible systems~\cite{QWEntanglement}. Controlled entanglement generation has currently a place at the forefront of research, as it is a fundamental resource in quantum computation and cryptography and therefore a pre-requisite for the construction of reliable devices for quantum information processing~\cite{entanglementgeneral}. The role of entanglement in different processes in nature is also currently under investigation~\cite{NatureEntanglement}.

Recently, an experimental realization of a linear quantum walk of a single neutral atom in a spin-dependent one-dimensional optical lattice has been reported, using site-resolved fluorescence imaging to demonstrate the spatial coherence of the final wave function~\cite{Karski:09}. Its conceptually straightforward generalization to higher-dimensions can however be experimentally demanding and would require significant technological efforts. {\it Here, we propose a novel scheme for a two-dimensional quantum walk that significantly reduces the resources necessary for its realization}. This is, in particular, true for an experimental setup similar to Ref.~\cite{Karski:09}; our proposal can however be easily adapted to other schemes which are able to realize one-dimensional quantum walks. In the remainder of the paper, we demonstrate that our scheme is able to mimic perfectly a specific spatial probability distribution associated with a well-known two-dimensional quantum walk. Moreover, it is efficient in the generation of spatial entanglement between the two orthogonal directions of the lattice on which the walker is moving.

To introduce our notation, let us start by describing a discrete-time quantum walk (we will not deal with continous-time quantum walks in this work) for the simpler one-dimensional model. In order to do that, we define a two-dimensional Hilbert space, $\cal{H}_C$ (coin space), spanned by $\{\miniket{0},\miniket{1}\}$, and an infinite dimensional Hilbert space, $\cal{H}_W$ (walker space), spanned by $\{\miniket{i}\}$, with $i$ assuming all possible integer values. The state of the system is described as a vector in $\cal{H}_C\otimes\cal{H}_W$ and the evolution of the system is given by a sequence of conditional shift and coin operations. We describe the conditional shift operation as
\begin{equation}
\hat{S}=\miniket{0}_C\minibra{0}\otimes\sum_{i\in \mathbb{Z}}\miniket{i-1}_W\minibra{i}+\miniket{1}_C\minibra{1}\otimes\sum_{i\in \mathbb{Z}}\miniket{i+1}_W\minibra{i}.
\end{equation}
If we consider the walker-component $\miniket{i}_W$ as describing the quantized position of the walker on a line, with increasing numbers from left to right, the effect of $\hat{S}$ is to move the walker one step to the left (right) when the coin-component is in the state $\miniket{0}_C$ ($\miniket{1}_C$). In the original quantum walk~\cite{Aharonov:93} the coin operation was chosen to be the Hadamard gate
\begin{equation}
\hat{H}=\frac{1}{\sqrt{2}}
\begin{pmatrix}
1&1\\
1&-1
\end{pmatrix}\;,
\end{equation}
but it can be any unitary operation acting on $\cal{H}_C$ only. A single time step consists here of a coin operation followed by a shift operation.

In the same way that classical random walks can be generalized to multi-dimensions, quantum walks do not have to be restricted to one dimension either and different two-dimensional situations have been analyzed in literature~\cite{2DQW}. However, the higher-dimensional walker space has required a similar increase in the dimension of the coin space in these schemes. In particular, for a two-dimensional walker, the coin system has been taken as four-dimensional, which can equivalently be interpreted as two different two-level coins or a single four-level coin. For a simpler illustration, we consider the second case. The states of the computational basis of the coin $\miniket{0}_C$, $\miniket{1}_C$, $\miniket{2}_C$ and $\miniket{3}_C$ then correspond to movements in the left-down, left-up, right-down and right-up directions, respectively. The specific choice of the coin operation distinguishes a number of different classes of two-dimensional quantum walks~\cite{2DQW}.

\begin{figure}[t]
\centerline{(a) \hskip4cm (b) \hskip0.375cm}
\centerline{\psfig{figure=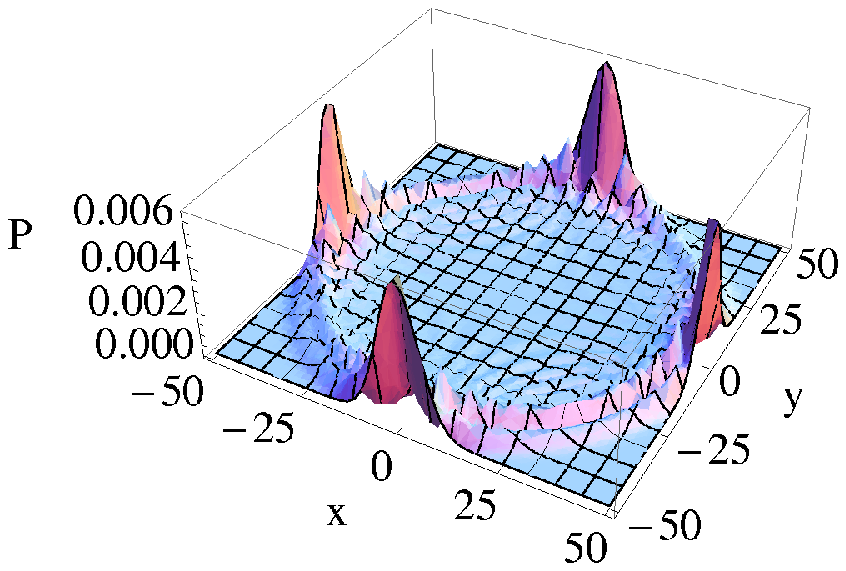,height=2.5cm}\hskip0.25cm\psfig{figure=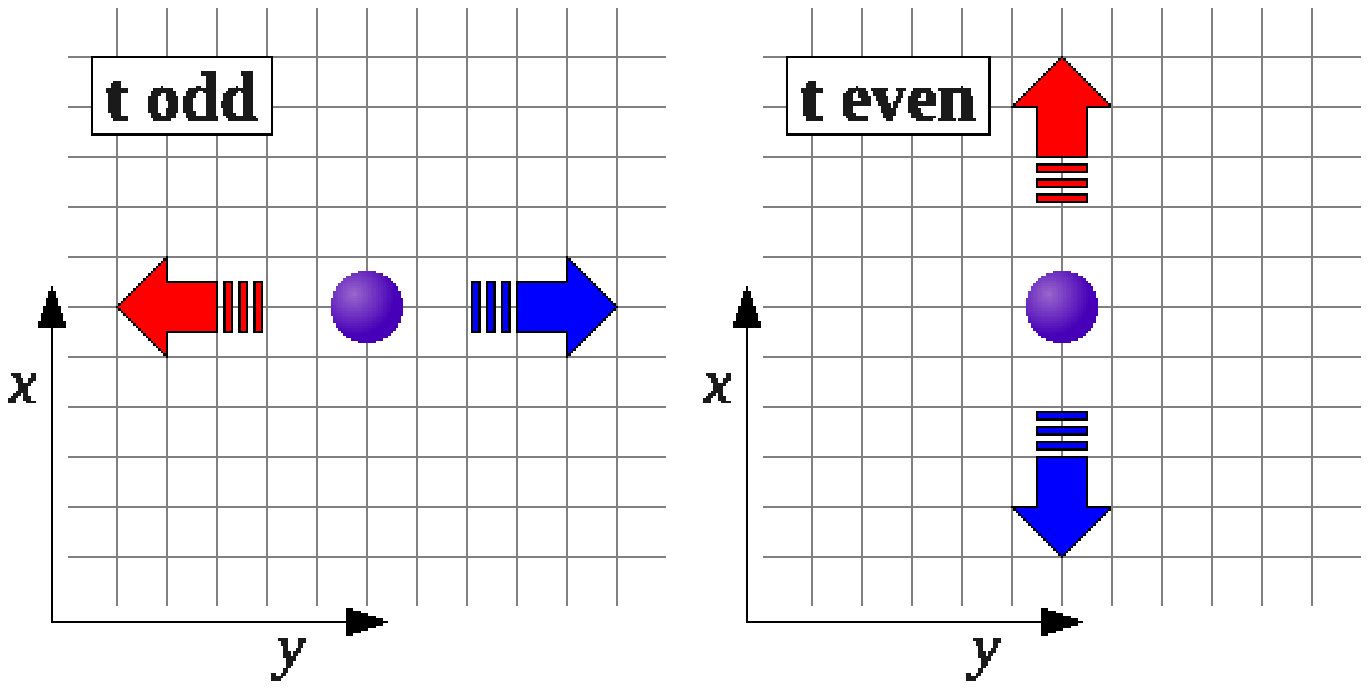,height=2.375cm}}
\caption{(a) Spatial probability distribution after $t=50$ steps of a two-dimensional Grover walk with the initial state of the coin as given in Eq.~\eqref{eq:GroverLocalized}. Only the sites with even $x$ and $y$ are shown, as the probability is zero for all odd sites. (b) Sketch of the proposed scheme, in line with the experiment in Ref.~\cite{Karski:09}. A two-dimensional spin-dependent optical lattice can alternately be shifted into the two orthogonal directions.}
\label{fig:GroverWalkNonLocalizedAndSketch}
\end{figure}

Let us focus our attention on the two-dimensional quantum walk known as the Grover walk. This particular walk has raised the interests of the scientific community, as it can be used in order to implement the two-dimensional Grover search algorithm~\cite{GroverSearch}. The corresponding coin operation is given by
\begin{equation}
\hat{G}=
\frac{1}{2}\left(
\begin{array}{rrrr}
-1&1&1&1\\
1&-1&1&1\\
1&1&-1&1\\
1&1&1&-1
\end{array}\right),
\end{equation}
and it has been shown that for this specific coin operation the walker is always localized ({\it i.e.}, the probability to find it at the origin is asymptotically larger than $0$ for $t\rightarrow\infty$), except if the coin is in the particular initial state~\cite{localization}
\begin{equation}
\frac{1}{2}(\miniket{0}_C-\miniket{1}_C-\miniket{2}_C+\miniket{3}_C).
\label{eq:GroverLocalized}
\end{equation}
The spatial probability distribution for this case can be obtained by tracing out the state of the coin and we show this distribution after $t=50$ time steps in
Fig.~\ref{fig:GroverWalkNonLocalizedAndSketch}(a).

{\it Unfortunately, these two-dimensional quantum walks are experimentally demanding}. The use of a four-level quantum system or two distinct qubits (which requires however an entangling gate at each time step) as the coin complicates the realization and only a few feasible physical implementations have been proposed so far (see, for instance, Ref.~\cite{Eckert:05}). In order to reduce the experimental challenges, we suggest considering a different two-dimensional quantum walk, in which the coin is a single qubit, and the movement on the $x$ and $y$ directions are alternate [according to the sequence: coin operation - movement on $x$ - coin operation - movement on $y$, see Fig.~\ref{fig:GroverWalkNonLocalizedAndSketch}(b)]. For this, we start with the initial state of the coin
\begin{equation}
\frac{1}{\sqrt{2}}(\miniket{0}_C+i\miniket{1}_C),
\label{eq:AlternateInitial}
\end{equation}
as in the original symmetric one-dimensional quantum walk, and consider the Hadamard gate as the coin operation. A time step consists here of a sequence of the two Hadamard operations and the two movement on the $x$ and $y$ directions. Surprisingly, if we analyze the spatial probability distribution of this two-dimensional quantum walk, we  find exactly the same result as for the non-localized Grover walk described above [see Fig.~\ref{fig:GroverWalkNonLocalizedAndSketch}(a)].

To prove formally this equivalence, in what follows we will show how the coefficients of the Grover walk in the non-localized case can be mapped to the coefficients of the alternate walk state with the initial condition of the coin given by Eq.~\eqref{eq:AlternateInitial}. For this, we define the basis states of the Grover walk as $\{\miniket{x,y,c}\}$, where $x,y\in\mathbb{Z}$ denote the position along the $x$ and $y$ directions, respectively, and $c\in\{0,1,2,3\}$ describes the state of the coin. At the same time, we decompose the state of our alternate walk over the basis $\{\miniket{x,y,c'}\}$, where $c'\in\{0,1\}$ is the state of the single-qubit coin. The coefficients in the decomposition of the Grover walk and of the alternate walk are then defined as $\alpha_{x,y,c}(t)$ and $\beta_{x,y,c'}(t)$, respectively. It is easy to note that, for the initial states under consideration, the $\alpha_{x,y,c}(t)$'s are real numbers, while the $\beta_{x,y,c'}(t)$'s are complex numbers. We will now show that
\begin{align}
\beta_{x,y,0}(t)&=(-1)^te^{i\pi/4}[\alpha_{x,y,0}(t)+i\alpha_{x,y,2}(t)],
\label{eq:beta1}\\
\beta_{x,y,1}(t)&=(-1)^te^{i\pi/4}[-\alpha_{x,y,1}(t)+i\alpha_{x,y,3}(t)].
\label{eq:beta2}
\end{align}
For this, let us first demonstrate that the amplitudes satisfy the properties
\begin{align}
\alpha_{x-1,y,0}(t)+\alpha_{x-1,y,1}(t)+\alpha_{x+1,y,2}(t)+\alpha_{x+1,y,3}(t)&=0,
\label{eq:alpha1}\\
\alpha_{x,y-1,0}(t)+\alpha_{x,y-1,2}(t)+\alpha_{x,y+1,1}(t)+\alpha_{x,y+1,3}(t)&=0,
\label{eq:alpha2}
\end{align}
if the walk starts at the origin with the initial state corresponding to
\begin{align}
\begin{split}
\label{eq:alpha0}
&\alpha_{0,0,0}(0)=1/2,\qquad\alpha_{0,0,1}(0)=-1/2,\\
&\alpha_{0,0,2}(0)=-1/2,\quad\;\alpha_{0,0,3}(0)=1/2
\end{split}
\end{align}
and all other $\alpha$'s at $t=0$ being zero.

Our proof works by induction on $t$. It is easy to verify, by means of a direct calculation, that Eqs.~\eqref{eq:alpha1} and \eqref{eq:alpha2} are satisfied at $t=0$. For this, we only need to evaluate them for the values $(x,y)$ equal to $(1,0)$, $(-1,0)$, $(0,1)$ and $(0,-1)$, as for all other values the $\alpha$'s are initially zero. Then, we assume that both Eqs.~\eqref{eq:alpha1} and \eqref{eq:alpha2} are true for any couple of values $(x,y)$ at some time $t$, and we need to prove that they hold at time $t+1$. Starting with the left hand side of Eq.~(\ref{eq:alpha1}), we have
\begin{align}
&\sum_{i=0}^1\alpha_{x-1,y,i}(t+1)+\sum_{i=2}^3\alpha_{x+1,y,i}(t+1)\nonumber\\
&=\sum_{j=0}^3\left[\sum_{i=0}^1G_{ij}\alpha_{x,y+(-1)^i,j}(t)+\sum_{i=2}^3G_{ij}\alpha_{x,y+(-1)^i,j}(t)\right]\nonumber\\
&=\sum_{j=0}^3\Big[(G_{0j}+G_{2j})\alpha_{x,y+1,j}(t)+(G_{1j}+G_{3j})\alpha_{x,y-1,j}(t)\Big]\nonumber\\
&=\sum_{i=0}^1\alpha_{x,y-1,2i}(t)+\sum_{i=0}^1\alpha_{x,y+1,2i+1}(t),
\end{align}
which is identically zero because we have assumed Eq.~\eqref{eq:alpha2} true at time $t$. Here, $G_{ij}$ $(i,j=0,1,2,3)$ is the element of the matrix $\hat{G}$ corresponding to $\miniket{i}_C\minibra{j}$. It is straightforward to proceed in the same way to prove Eq.~\eqref{eq:alpha2} at time $t+1$, assuming that Eq.~\eqref{eq:alpha1} is true at time $t$.

We can now prove the relations in Eqs.~\eqref{eq:beta1} and \eqref{eq:beta2} for the initial conditions for the alternate walk given by
\begin{equation}
\beta_{0,0,0}(0)=1/\sqrt{2},\qquad\beta_{0,0,1}(0)=i/\sqrt{2},
\label{eq:beta0}
\end{equation}
and all other $\beta$'s being zero at $t=0$ and Eq.~\eqref{eq:alpha0} for the Grover walk. Again, we proceed by induction in $t$. At $t=0$, all the amplitudes $\alpha$'s and $\beta$'s are zero outside of the origin $(x,y)=(0,0)$ and using Eqs.~\eqref{eq:alpha0} and \eqref{eq:beta0} one can straightforwardly verify that Eqs.~\eqref{eq:beta1} and \eqref{eq:beta2} are fulfilled at the origin. We then assume that Eqs.~\eqref{eq:beta1} and \eqref{eq:beta2} are true [for any couple of values $(x,y)$] at some time $t$, and we prove that they hold for the time $t+1$. For our alternate walk this leads to
\begin{align}
&\beta_{x,y,i}(t+1)=\frac{1}{2}\big\{\beta_{x+1,y+(-1)^i,0}(t)+\beta_{x+1,y+(-1)^i,1}(t)\nonumber\\
&+(-1)^i\big[\beta_{x-1,y+(-1)^i,0}(t)-\beta_{x-1,y+(-1)^i,1}(t)\big]\big\},
\label{eq:alt}
\end{align}
with $i=0,1$, while, for the Grover walk, we find
\begin{equation}
\alpha_{x,y,k}(t+1)=\sum_{j=0}^3G_{kj}\alpha_{x+(-1)^m,y+(-1)^n,j}(t),
\end{equation}
with $k=0,1,2,3$, $m={\rm Int}[k/2]$ and $n={\rm mod}[2,k]$. Starting from Eq.~\eqref{eq:alt} with $i=0$ and assuming Eq.~\eqref{eq:beta1} true at time $t$, we can therefore write
\begin{align}
&\beta_{x,y,0}(t+1)\nonumber\\
&=\frac{1}{2}(-1)^te^{i\frac{\pi}{4}}\Big\{\sum_{i=0}^1(-1)^i\alpha_{x+1,y+1,i}(t)+\sum_{i=0}^1\alpha_{x-1,y+1,i}(t)\nonumber\\
&+i\big[\sum_{i=2}^3\alpha_{x+1,y+1,i}(t)+\sum_{i=2}^3(-1)^i\alpha_{x-1,y+1,i}(t)\Big]\Big\}.
\label{eq:ind1}
\end{align} 
Now, by using Eq.~\eqref{eq:alpha1}, straightforward calculations lead to
\begin{align}
&\beta_{x,y,0}(t+1)\nonumber\\
&=(-1)^{t+1}e^{i\frac{\pi}{4}}\sum_{j=0}^3\Big[ 
G_{0j}\alpha_{x+1,y+1,j}(t)+iG_{2j}\alpha_{x-1,y+1,j}(t)\Big]\nonumber\\
&=(-1)^{t+1}e^{i\pi/4}\big[\alpha_{x,y,0}(t+1)+i\alpha_{x,y,2}(t+1)\big],
\end{align}
which completes the proof for Eq.~\eqref{eq:beta1}. An analogous analysis allows us to prove the partner Eq.~\eqref{eq:beta2}, showing that the alternate walk and the Grover walk generate the same spatial density distribution for the aformentioned initial conditions.

\begin{figure}[t]
\centerline{(a) \hskip4cm (b)}
\centerline{\psfig{figure=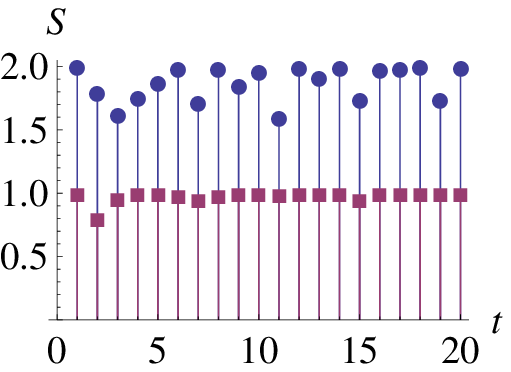,height=3cm}\hskip0.5cm\psfig{figure=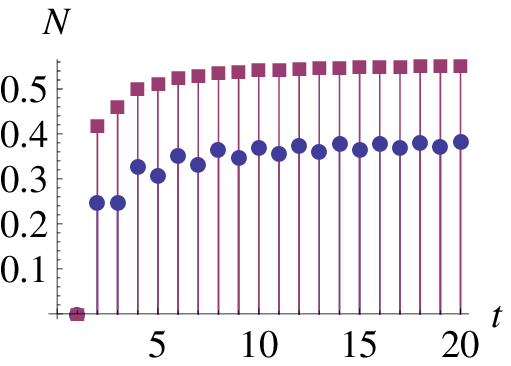,height=3cm}}
\caption{(a) Coin-position entanglement against the number of steps $t$ for the Grover walk (blue circles) with the initial conditions as in Eq.~\eqref{eq:alpha0} and for our alternate walk (purple squares) with the initial conditions as in Eq.~\eqref{eq:beta0}. (b) Entanglement between $x$ and $y$ position of the particle for the Grover walk and our alternate walk, with the same conditions as in panel (a).}
\label{fig:entanglement}
\end{figure}

As already stated, one interesting aspect of quantum walks is their ability to generate entanglement. Even if the spatial probability distributions are the same in the two different walks investigated in this paper, the final state after a fixed number of steps is clearly different. It is therefore interesting to make a comparison between them in terms of entanglement generation. A well investigated feature in one- as well as two-dimensional quantum walks is the generation of entanglement between the state of the coin and the position of the walker (coin-position entanglement)~\cite{cpentanglement}. Since the evolution of the whole system under the action of the walk is unitary, the total state remains pure, and it is possible to use the von Neumann entropy of the reduced density matrices to measure the coin-position entanglement. For this, we need to evaluate $S(\rho_C)=S(\rho_W)$, where $\rho_C$ ($\rho_W$) is the reduced density matrix of the coin (walker). To make a comparison between the two different dynamical walks described above, we first calculate $S(\rho_C)$ for the Grover walk, with the initial conditions as in Eq.~\eqref{eq:alpha0}. The results for $t=20$ steps are shown in Fig.~\ref{fig:entanglement}(a). One can clearly see that the entropy is quite high throughout the process and close to its maximum, which in this case is $S=\log_24=2$, because the Hilbert space of the coin is four-dimensional. The coin-position entanglement in the alternate walk, with the initial conditions as in Eq.~\eqref{eq:beta0}, is also shown in the same figure. Similarly to the Grover walk, the entropy quickly reaches values close to its maximum (the fluctuations are less significant here). However, due to the fact that the Hilbert space of the coin is only two-dimensional, this corresponds to $S=\log_22=1$.

Let us now consider a different kind of entanglement. Suppose that we want to entangle the orthogonal directions of the lattice on which the walker is moving (we will denote this as x-y spatial entanglement). In order to do so, we need to trace out the degree of freedom embodied by the coin.  This can be an advantage of our alternate walk in terms of the amount of entanglement generated, due to the smaller dimension of the coin space. However, since the state after tracing out the coin is mixed, the von Neumann entropy can no longer be used as a measure of entanglement and we will, for this reason, use the negativity $N$ of the partial transpose, in its generalization for higher-dimensional systems (so as to have $0\le N\le 1$)~\cite{quditsnegativity}. The results for both walks for a number of steps up to $t=20$ are shown in Fig.~\ref{fig:entanglement}(b), and one can clearly see that for the alternate walk the x-y spatial entanglement is, in fact, always larger.

Let us finally briefly discuss a possible experimental implementation of our proposed walk, following Ref.~\cite{Karski:09}. In the experiment of Karski {\sl et al.}, a single cesium atom is able to {\sl walk} in a one-dimensional optical lattice in position space. The conditional shift operator is realized by continuously controlling the trap polarization in a way so as to move the spin state $\miniket{0}$ ($\miniket{1}$) to the right (left). Coin operations are obtained by proper laser pulses and the atom distribution is probed by fluorescence imaging. The setting can be adapted to our walk by keeping the same coin operation and allowing the optical lattice (in this case, a two-dimensional one) to alternately shift into the two orthogonal directions, as sketched in Fig.~\ref{fig:GroverWalkNonLocalizedAndSketch}(b). {\it In terms of experimental challenges, this scheme is therefore more feasible and could in principle be realized with the state-of-the-art current technology}. In fact, it just exploits a straightforward extension of the existing setup and is thus less demanding than performing a coin operation on four different internal states of the particle and moving them into four possible directions simultaneously (as required for an implementation of the original Grover walk).

In summary, we have shown that a two-dimensional quantum walk with a four-dimensional coin space can be equivalent to two one-dimensional walks in alternating directions at alternating time steps, for which only a two-dimensional coin space is required. In particular, we have formally shown that such a walk is able to perfectly mimic the spatial probability distribution of the well-known Grover walk. We have analyzed the behavior of two different kinds of entanglement for both walks and found that the presented alternate walk is more efficient at generating spatial x-y entanglement. Finally, we have sketched a possible physical implementation of the scheme, which is in lines with the current experimental technology. A deeper investigation of two-dimensional quantum walks with single-qubit coins will be an exciting and interesting extension of the work presented here, due to the importance of quantum walks in the implementation of quantum algorithms (for instance, an extension to a more general coin operation, along the lines of Ref.~\cite{watabe}, is currently under investigation.). We believe that our results contribute to this task in a significant way.

We thank N. Lo Gullo and M. Paternostro for discussions. This work was supported by Science Foundation Ireland under grant no. 05/IN/I852 and 05/IN/I852 NS. C.D.F. is supported by the Irish Research Council for Science, Engineering and Technology.

\end{document}